\documentclass[aps,prl,twocolumn,groupedaddress]{revtex4-1}
\usepackage{calrsfs}
\usepackage{mathrsfs}
\usepackage{graphicx}
\usepackage{dcolumn}
\usepackage[utf8]{inputenc}
\usepackage[T1]{fontenc}
\usepackage{textcomp}
\usepackage{gensymb}
\usepackage{bm}
\usepackage{hyperref}
\usepackage{xcolor}
\hypersetup{colorlinks=true,linkcolor=red}
\usepackage{color}
\usepackage{fancyhdr}
\usepackage{url}
\usepackage{amssymb,amsfonts,amsmath,amsbsy,bm,t1enc,lipsum,latexsym}
\usepackage{changes}

\begin{document}

\normalem

\title{Photonic Damascene Process for Integrated High-Q Microresonator Based Nonlinear Photonics}

\author{M. H. P. Pfeiffer}
\affiliation{{\'E}cole Polytechnique F{\'e}d{\'e}rale de Lausanne (EPFL), CH-1015 Lausanne, Switzerland}

\author{A. Kordts}
\affiliation{{\'E}cole Polytechnique F{\'e}d{\'e}rale de Lausanne (EPFL), CH-1015 Lausanne, Switzerland}

\author{V. Brasch}
\affiliation{{\'E}cole Polytechnique F{\'e}d{\'e}rale de Lausanne (EPFL), CH-1015 Lausanne, Switzerland}

\author{M. Zervas}
\affiliation{{\'E}cole Polytechnique F{\'e}d{\'e}rale de Lausanne (EPFL), CH-1015 Lausanne, Switzerland}

\author{M. Geiselmann}
\affiliation{{\'E}cole Polytechnique F{\'e}d{\'e}rale de Lausanne (EPFL), CH-1015 Lausanne, Switzerland}

\author{J. D. Jost}
\affiliation{{\'E}cole Polytechnique F{\'e}d{\'e}rale de Lausanne (EPFL), CH-1015 Lausanne, Switzerland}

\author{T. J. Kippenberg}
\email[E-mail: ]{tobias.kippenberg@epfl.ch}
\affiliation{{\'E}cole Polytechnique F{\'e}d{\'e}rale de Lausanne (EPFL), CH-1015 Lausanne, Switzerland}

\begin{abstract}
High confinement, integrated silicon nitride (SiN) waveguides have recently emerged as attractive platform for on-chip nonlinear optical devices. The fabrication of high-Q SiN microresonators with anomalous group velocity dispersion (GVD) has enabled broadband nonlinear optical frequency comb generation. Such frequency combs have been successfully applied in coherent communication and ultrashort pulse generation. However, the reliable fabrication of high confinement waveguides from stoichiometric, high stress SiN remains challenging. Here we present a novel photonic Damascene fabrication process enabling the use of substrate topography for stress control and thin film crack prevention. With close to unity sample yield we fabricate microresonators with $1.35\,\mu\mathrm{m}$ thick waveguides and optical Q factors of $3.7\times10^{6}$ and demonstrate single temporal dissipative Kerr soliton (DKS) based coherent optical frequency comb generation. Our newly developed process is interesting also for other material platforms, photonic integration and mid infrared Kerr comb generation.
\end{abstract}

\maketitle

\section{Introduction}

Integrated silicon nitride (SiN) waveguides and resonators are an attractive platform for nonlinear optics \citep{Moss2013}. SiN waveguides combine the material's large bandgap (\textasciitilde 5 eV) and wide transparency range with CMOS fabrication methods and a large effective nonlinearity. Upon launching a femtosecond laser pulse inside a SiN waveguide the interplay of anomalous group velocity dispersion (GVD) and high effective nonlinearity ($\gamma\approx1.4\:\mathrm{W^{-1}m^{-1}}$) leads to efficient supercontinuum generation \citep{Halir2012}. Harnessing four-wave-mixing processes inside SiN waveguides low noise optical amplification of weak signals can be achieved \citep{Agha2012}. Moreover, the fabrication of high-Q SiN microresonators with anomalous GVD at telecom wavelengths has allowed for the observation of parametric oscillations and optical frequency comb generation in integrated SiN microresonators \citep{Levy2009}, based on the Kerr frequency comb generation mechanism first reported in 2007 \citep{DelHaye2007}. The spectral bandwidth of the generated frequency combs can attain an octave\citep{Okawachi2011}. 

Following this work, a significant advance in SiN nonlinear photonics was the generation of low noise frequency combs \citep{Herr2012,Saha:13}. Due to the comparatively large ratio of cavity linewidth to dispersion parameter of SiN microresonators, the nonlinear comb formation process takes place via the formation of subcombs that can lead to the loss of spectral coherence and high comb noise. It was shown that low phase noise comb states can still be attained via synchronization of the generated subcombs via $\delta-\Delta$ matching \citep{Herr2012,Papp:13} and have also been observed to allow ultrashort optical pulse generation \citep{Saha:13}. Using such phase locked comb states the high application potential of SiN microresonator based frequency combs was demonstrated in experiments on Tb/s coherent communication \citep{Pfeifle2014} and ultrafast optical waveform generation \citep{Ferdous2011}. Most recently, it has been demonstrated that temporal dissipative Kerr soliton (DKS) formation and DKS induced Cherenkov radiation are accessible in SiN microresonators \citep{Brasch2014}. This allows generation of frequency combs that are deterministically fully coherent, numerically predictable \citep{Coen2013,Herr2013a}, broadband (2/3 of an octave) and which exihibit a smooth spectral envelope, underscoring the potential of SiN based microresonators to serve as compact, mass producible, on-chip optical frequency comb generators.

\begin{figure}
	\includegraphics[width=1\columnwidth]{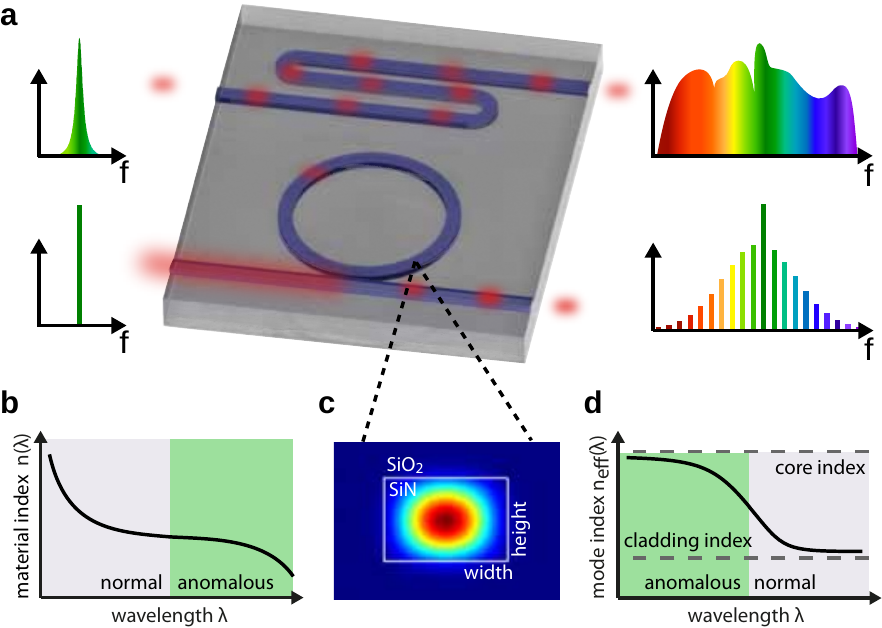}
	
	\protect\caption{\textbf{Nonlinear wavelength conversion in integrated waveguides and
			principles of their dispersion engineering.} (a) Upon launching a
		short light pulse into the waveguide (top) its initial spectrum is
		nonlinearly broadened. Coupling the light of a single frequency, continuous
		wave laser into the ring resonator (bottom) produces short soliton
		pulses with broad spectral envelope. (b) Most materials exhibit anomalous
		GVD towards longer wavelengths but not necessarily at telecom wavelengths.
		However, for efficient nonlinear wavelength conversion the GVD of
		the highly confining waveguides must be anomalous. (c), (d) By tailoring
		the waveguide cross section to highly confine the optical mode, the
		anomalous waveguide dispersion can overcompensate the normal material
		dispersion. Thus dispersion engineering determines the required waveguide
		dimensions for a given operating wavelength.\label{fig:Waveguide-dispersion-engineering}}
\end{figure}

The above advances relied on engineering of the waveguide's GVD through tailoring the dimensions of the waveguide cross section. The GVD for a given waveguide cross section can be approximated as the sum of the material GVD (Figure \ref{fig:Waveguide-dispersion-engineering}b) and its waveguide GVD (Figure \ref{fig:Waveguide-dispersion-engineering}d). While the material GVD is typically normal at short wavelengths and anomalous at longer wavelengths, the waveguide GVD exhibits the opposite behaviour. Thus by adjusting the waveguide cross section the waveguide dispersion can be tailored to overcompensate the material dispersion. At telecom wavelengths the material GVD of SiN is normal but by employing a high confinement waveguide with height in excess of $0.7\mu\mathrm{m}$ an effective anomalous GVD is tailored. Consequently even larger waveguide cross sections are needed in order to enable SiN based nonlinear integrated optics, especially Kerr comb generation, in the mid infrared ($3-5\,\mu\mathrm{m}$) \citep{Griffith2015}.

\section{Problems of subtractive waveguide fabrication}
\label{sec:Problems-of-subtractive}

\begin{figure}
	\includegraphics[width=1\columnwidth]{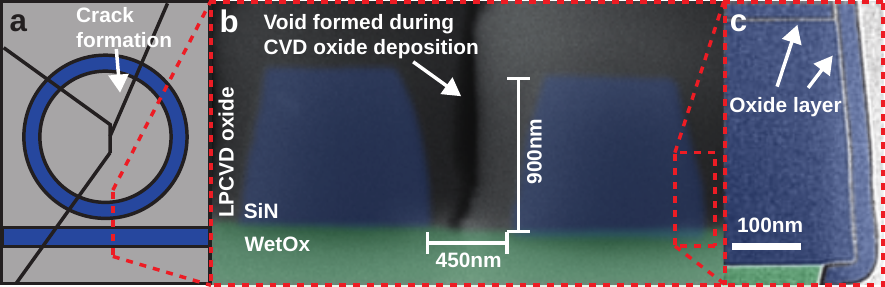}
	
	\protect\caption{\textbf{Problems in subtractive SiN waveguide fabrication.} (a) Schematic
		top view of a microresonator (blue) with cracks formed by the highly
		stressed SiN film. (b) Cleaved waveguide cross section showing the
		limited aspect ratio between adjacent waveguides (a geometry used
		for resonator waveguide coupling) and void formation of the low temperature
		oxide (LTO) cladding layer. (c) Transmission electron microscope (TEM)
		image showing the thin oxide layers formed during multi-step deposition
		of SiN.\label{fig:Schematic-process-flow}}
\end{figure}

The fabrication of such high confinement waveguides with heights larger than $0.7\mu\mathrm{m}$ from stoichiometric SiN is challenging. So far, fabrication typically relies on a subtractive process approach: the waveguide structures are etched into a previously deposited thin film. The fabrication challenges posed by this process approach are summarized in Figure \ref{fig:Schematic-process-flow}a-c.

While plasma enhanced chemical vapor deposition (PECVD) allows for crack free SiN thin film deposition, high-Q optical microresonators were only achieved based on low pressure chemical vapor deposited (LPCVD), stoichiometric SiN thin films. The high temperatures during film deposition ($\sim\mathrm{770\text{\textdegree}C}$) and for subsequent annealing (up to $1200\text{\textdegree}\mathrm{C}$) have been shown to lead to low material absorption losses \citep{Shaw2005}. But they also lead to highly tensile stressed films (typically $>800\,\mathrm{MPa}$), that are prone to crack formation (Figure \ref{fig:Schematic-process-flow}a). Such cracks cause high scattering losses in the waveguide and are thus unacceptable for high-Q microresonators. So far only a multi-step growth approach, that partially relaxes film stress through thermic cycling during deposition, can reduce the crack density per wafer to an acceptable level \citep{Gondarenko2009}. Additional crack stop structures have been introduced as the device yield was still reduced by the remaining cracks \citep{Luke2013}.

Furthermore, the gap between two closely spaced, thick waveguides can present a critical aspect ratio for several processing steps. The dry etch process that defines the waveguide structures often has a limited anisotropy and aspect ratio dependent etch rates (ARDE). It is typically optimized to produce smooth waveguide sidewalls, to limit scattering losses, but often does not accurately transfer the resist mask pattern. Figure \ref{fig:Schematic-process-flow}b shows an example: the measured waveguide gap of $450\,\mathrm{nm}$ deviates strongly from the design value of $200\,\mathrm{nm}$. This is due to the limited anisotropy of the etch process applied which is primarily designed for low sidewall roughness. Moreover voids can form in between closely spaced waveguides during the $\mathrm{SiO_{2}}$ cladding deposition, due to the limited conformality of the deposition processes.

Finally, it was found that nanometer thin silicon oxide layers form between the individual SiN layers when using multi-step deposition with thermal cycling (Figure \ref{fig:Schematic-process-flow}c), that may be undesirable in certain applications.

\section{The photonic Damascene process}
\label{sec:The-photonic-Damascene}
\begin{figure}
	\includegraphics[width=1\columnwidth]{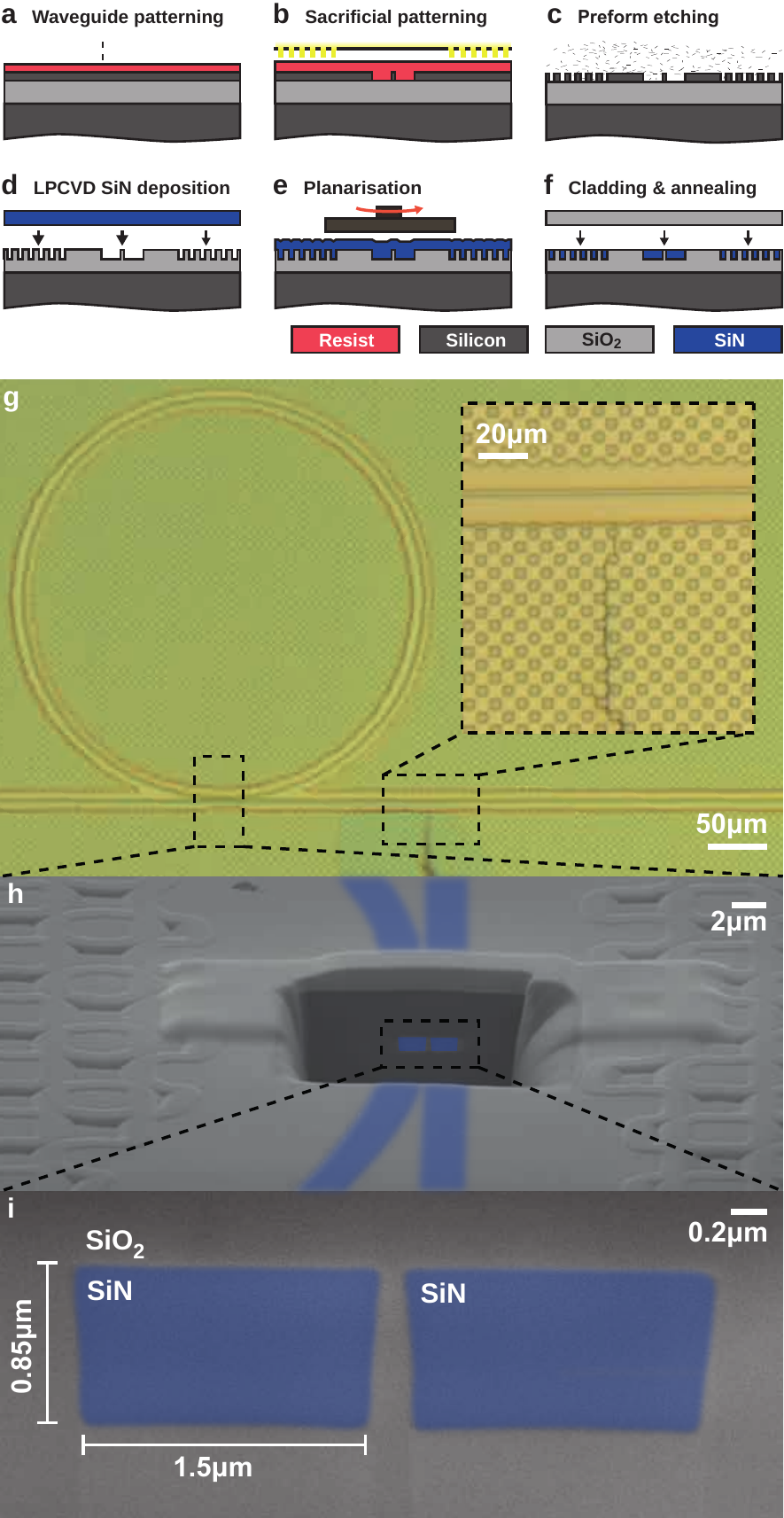}
	
	\protect\caption{\textbf{Photonic Damascene process for integrated SiN waveguides.}
		(a)-(f) Schematic process flow of the photonic Damascene process.
		(g) Optical image of a SiN microresonator surrounded by the stress
		release structure (rectangle dimensions $5\,\mu\mathrm{m}\times 5\,\mu\mathrm{m}$).
		The zoomed inset reveals the $10\,\mu\mathrm{m}$ wide area to each
		side of the waveguide that has no stress release structure to avoid
		scattering losses. The crack formed due to incomplete removal of excess
		SiN but does not penetrate the waveguide. (h),(i) Focused ion beam
		(FIB) cross section of the coupling region between the ring resonator
		and the bus waveguide, revealing a waveguide resonator separation
		below 200 nm. The individual SiN waveguides (blue) are $1.5\,\mu\mathrm{m}$
		wide and $0.85\,\mu\mathrm{m}$ high and homogenously filled with
		SiN. The coupling region is free of voids and no effect of the waveguide
		proximity on the waveguide shape is observed.\label{fig:Focused-ion-beam}}
\end{figure}

Inspired by the additive patterning process (commonly referred to as "Damascene process'' or "dual-Damascene process'') for microprocessor copper interconnects \citep{Kaanta1991}, we present here a novel "photonic Damascene process'' that overcomes the aforementioned challenges. The process relies on substrate prepatterning prior to core material deposition and a subsequent planarization step (Figure \ref{fig:Focused-ion-beam}a-f). Using this process we are able to reliably fabricate for the first time SiN waveguides, as well as high-Q microresonators, with unprecedentedly large waveguide dimensions, due to a novel stress control technique based on dense substrate prepatterning. Previously this process approach has been demonstrated for waveguides based on amorphous silicon \citep{Sun2009} and recently also for limited height SiN waveguides \citep{Epping2015}. 

In a first step the waveguide pattern is defined using electron beam lithography (Figure \ref{fig:Focused-ion-beam}a). The resist mask is transferred via dry etching into an amorphous silicon (a-Si) hardmask. Then the dense stress release pattern is defined using photolithography and transferred into the hardmask layer as well (Figure \ref{fig:Focused-ion-beam}b). During the subsequent preform etch the a-Si hardmask is thus structured with both patterns which are dry etched into the wet thermal oxide layer of the substrate (Figure \ref{fig:Focused-ion-beam}c). The 10:1 etch selectivity between the a-Si hardmask and the thermal oxide avoids geometry limitations and sidewall roughness due to mask erosion. Before the SiN thin film can be deposited the a-Si hardmask layer is stripped. Due to the stress release effect of the densely prestructured substrate the LPCVD SiN thin film can now be deposited in one run up to the desired thickness (Figure \ref{fig:Focused-ion-beam}d). The excess SiN material is removed using chemical mechanical planarization (CMP) and a polished substrate exposing only the waveguides top surface is obtained (Figure \ref{fig:Focused-ion-beam}e). Next, thermal annealing is used to densify the SiN film, reducing its content of residual hydrogen. Finally, a cladding layer of low temperature oxide (LTO) is deposited and the chip side facets are dry etched before separating the wafer into individual chips (Figure \ref{fig:Focused-ion-beam}f). 

Figure \ref{fig:Focused-ion-beam}g shows an optical image of a ring resonator fabricated with the photonic Damascene process, as well as a focused ion beam (FIB) cross section of the coupling region. The waveguides are surrounded by a dense pattern which relaxes the SiN film stress and thus ensures high yield fabrication. The FIB cross section (Figure \ref{fig:Focused-ion-beam}h,i) reveals SiN waveguides with $1.5\,\mu\mathrm{m}$ width and $0.85\,\mu\mathrm{m}$ height. The waveguides have nearly vertical sidewalls and a flat top surface, indicating that the planarization was successful and no top surface topography was transferred into the waveguide. The deposited SiN uniformly fills the preform and no boundaries are observed within the SiN core.

One of the inherent advantages of the process is its ability to produce closely spaced waveguides comprising narrow aspect ratios between them. As seen in Figure \ref{fig:Focused-ion-beam}i the aspect ratio of the coupling gap between resonator and bus waveguide is better than 4:1 (with a gap below 200 nm) while being free of voids or proximity related effects, which is challenging to achieve in subtractive processing.

\section{Stress control by dense Prestructuring}
\label{sec:Stress-release-by}
Furthermore, the photonic Damascene process approach allows the elegant use of substrate topography for stress control and crack prevention in the deposited thin film. Crack formation in high stress LPCVD SiN thin films is a long standing problem that has so far often limited film thickness to at most $250\,\mathrm{nm}$ \citep{Daldosso2004}. However, as shown in reference \citep{Nam2012}, crack development in SiN thin films strongly depends on the substrate topography. Here we show that the high tensile stress of micron thick, stoichiometric LPCVD SiN films can be efficiently relaxed through dense prestructuring of the substrate, allowing for crack free high confinement waveguide processing with high yield.

The dense prestructuring consists of introducing a checkerboard structure (rectangle dimensions $5\,\mu\mathrm{m}\times 5\,\mu\mathrm{m}$) around the waveguides (Figure \ref{fig:Focused-ion-beam}g). The checkerboard covers the full wafer but the waveguides are surrounded by a $10\,\mu\mathrm{m}$ wide checkerboard free area to prevent scattering losses. While the pattern used here is a simple checkerboard the stress release effect is not strongly dependent on the actual pattern. It was found that the stress control originates mainly from the topography modulation which prevents the formation of a continous thin film. 

After deposition no cracks are observed if the deposited film thickness is not exceeding the preetched depth by more than $100\,\mathrm{nm}$. Importantly the dense prepatterning prevents crack formation also during the chemical mechanical planarization. Additionally no wafer bending, due to asymmetric wafer coating with highly stressed thin films, is observed. Yet, if the planarization step does not remove all excess SiN, cracks can still occur during the post deposition processing, as seen in the inset of Figure \ref{fig:Focused-ion-beam}g. Advantageously they do not penetrate the optical waveguide but are stopped at the pattern-free area around the waveguide. 

\begin{figure}
	\includegraphics[width=1\columnwidth]{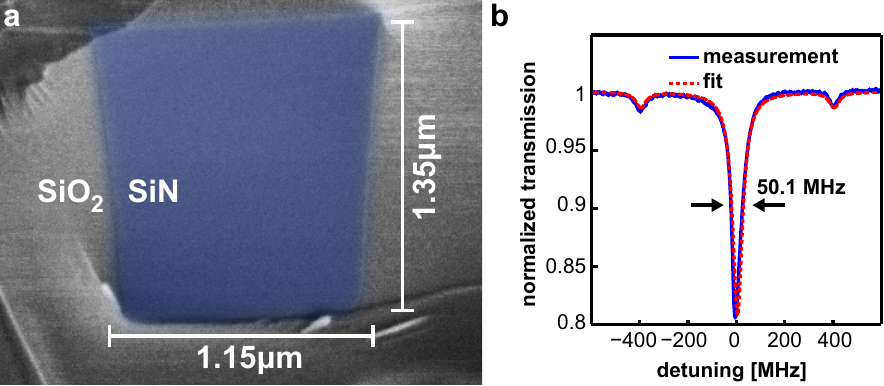}
	
	\protect\caption{\textbf{Example of thick, high-Q SiN resonators.} (a) Cleaved cross
		section through a SiN waveguide (blue) with $1.35\,\mu\mathrm{m}$
		height and $1.15\,\mu\mathrm{m}$ width fabricated using the photonic
		Damascene process. (b) Resonance linewidth at $\lambda=1598\mathrm{\, nm}$
		of a microresonator with a radius of $238\,\mu\mathrm{m}$ ($\mathrm{FSR}\approx100\mathrm{\, GHz}$)
		built from $1.35\,\mu\mathrm{m}$ high and $1.5\,\mu\mathrm{m}$
		wide waveguides. A resonance linewidth of $50\,\mathrm{MHz}$ is extracted
		which corresponds to a quality factor of $\mathrm{Q}=3.7\times10^{6}$.
		The sidebands of the $400\,\mathrm{MHz}$ calibration tone are visible.\label{fig:best_performance}}
\end{figure}

Using the dense prepatterning we deposit and process $1.2\,\mu\mathrm{m}$ thick, high stress SiN thin films, deposited without thermic cycling, and achieve a sample yield of more than 95\%. Figure \ref{fig:best_performance}a shows a cleaved cross section of a resulting waveguide with a record thickness of $1.35\,\mu\mathrm{m}$.

\section{Device characterization}
\label{sec:Device-characterization}
We characterize microresonator devices fabricated using the photonic Damascene process with respect to their loss and dispersion. Figure \ref{fig:best_performance}b shows the linewidth of a resonator mode at $\lambda=1598\,\mathrm{nm}$. The resonator has a radius of $238\,\mu\mathrm{m}$ (free spectral range: $\mathrm{FSR}=\mathrm{D_{1}}/2\pi\approx100\mathrm{\, GHz}$) and its waveguide is $1.5\,\mu\mathrm{m}$ wide and $1.35\,\mu\mathrm{m}$ high. The resonance is strongly undercoupled so that the total linewidth $\kappa=\kappa_{0}+\kappa_{\mathrm{ex}}$ is dominated by the internal loss rate $\kappa_{0}$ and the external coupling rate $\kappa_{\mathrm{ex}}$ is small. An electro-optic modulator produced $400\,\mathrm{MHz}$ sidebands around the central resonance dip used to frequency calibrate the laser scan. By fitting the central transmission dip with a Lorentzian a linewidth of $\kappa/2\pi=50\mathrm{\, MHz}$ is extracted which corresponds to a quality factor of $\mathrm{Q}=3.7\times10^{6}$.

Anomalous GVD and an undistorted mode structure are essential for the nonlinear performance of integrated microresonators, in particular the generation of frequency combs using DKS \citep{Herr2013}. Here we use frequency comb assisted diode laser spectroscopy \citep{DelHaye2009} to assess dispersion and mode spectrum of a resonator with $238.2\,\mu\mathrm{m}$ radius, $1.75\,\mu\mathrm{m}$ waveguide width and $0.85\,\mu\mathrm{m}$ waveguide height.

\begin{figure}
	\includegraphics[width=1\columnwidth]{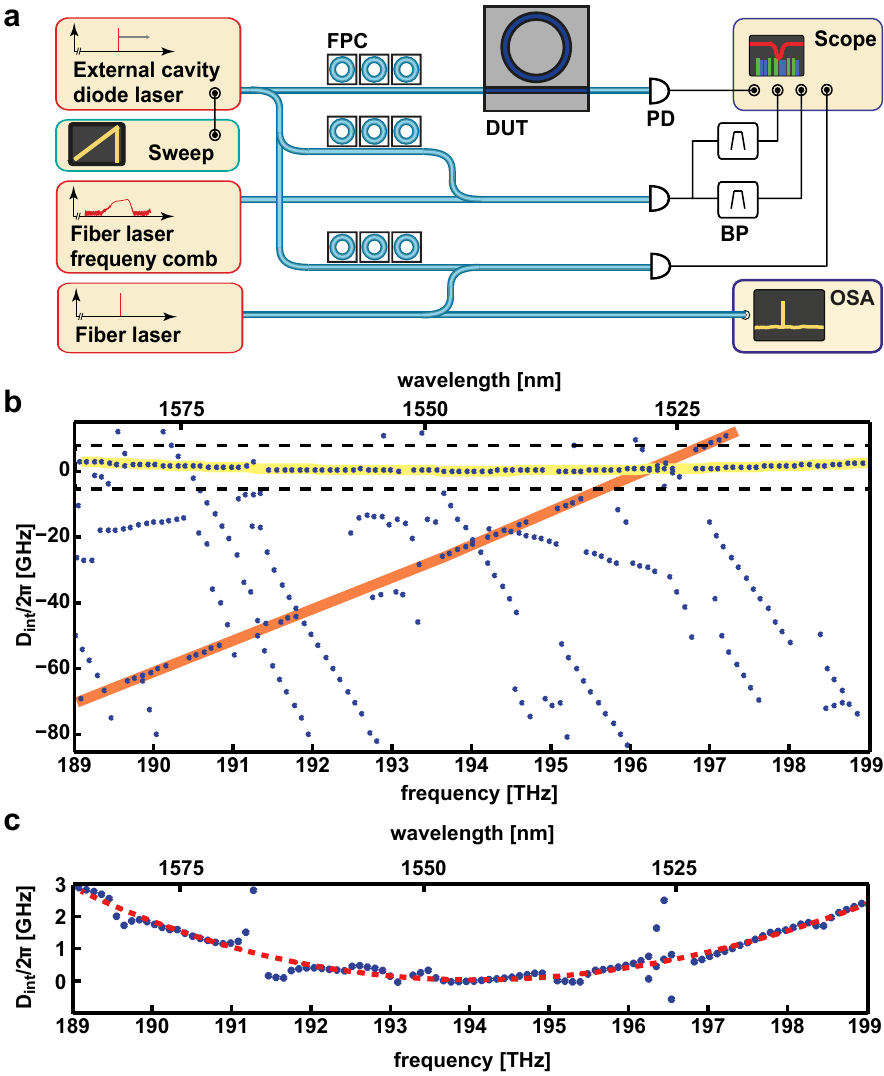}
	
	\protect\caption{\textbf{Dispersion characterization of microresonators.} (a) Schematic
		of the setup (adapted from \citep{Riemensberger2012}) used to record
		a frequency calibrated transmission amplitude of the device under
		test (DUT). The bandpass filters (BP) transform the beat signal between
		scanning ECDL and frequency comb respectively the fiber laser into
		marker peaks for relative respectively absolute frequency calibration.
		(b) Mode structure of a microresonator with $238.2\,\mu\mathrm{m}$
		radius, $1.75\,\mu\mathrm{m}$ waveguide width and $0.85\,\mu\mathrm{m}$
		waveguide height. The plot shows the deviation of each detected mode
		from a FSR of $95.65\mathrm{\, GHz}$. The $E_{11}^{x}$ mode family
		is underlaid in orange and the $E_{11}^{y}$ mode family in yellow.
		(c) Magnified view of the local parabolic dispersion of the $E_{11}^{y}$
		mode family. The detected resonances are fitted with a parabola (red
		dashed line) and a value of $\mathrm{D_{2}}/2\pi=0.49\mathrm{\, MHz}$
		is extracted (anomalous GVD).\label{fig:dispersionCharacterization}}
\end{figure}

Figure \ref{fig:dispersionCharacterization}a shows schematically the measurement setup used. The light of an external cavity diode laser (ECDL) is coupled into the device under test (DUT) using lensed fibers. By scanning the laser over $10\,\mathrm{THz}$ the device's transmission amplitude is recorded using photodiodes (PD). The frequency calibration of the laser scan is achieved by simultaneously recording the beat signal of the ECDL and the teeth of a fiber laser frequency comb. The beat signal is filtered by two bandpass filters that produce reference marker peaks. For wavelength calibration the beat signal of the ECDL with a stable, narrow linewidth fiber laser is recorded, the wavelength of which is measured with an optical spectrum analyzer (OSA). 

Due to the waveguide and material dispersion the resonances of the microresonator are not equidistantly spaced, and the free spectral range (FSR) changes with frequency. This can be expressed as an integrated deviation $\mathrm{D_{int}}$ of the resonance frequency from an equidistant grid defined by the central FSR: $D_{\mathrm{int}}(\mu)\equiv\omega_{\mu}-(\omega_{0}+D_{1}\mu)=D_{2}\frac{\mu^{2}}{2!}+D_{3}\frac{\mu^{3}}{3!}+\dots$, where $\mu\in\mathbb{Z}$ is the relative mode number counted from the central mode $\mu=0$, $D_{1}$ is the FSR of the central mode, $D_{2}$ and $D_{3}$ are the second and third order dispersion coefficients.

Figure \ref{fig:dispersionCharacterization}b shows an Echelle plot representation of the device's mode spectrum. The plot shows every identified resonance in the transmission trace as one point and plots their individual deviation from a central FSR of $95.65\,\mathrm{GHz}$. Resonances belonging to the same mode family are lined up, showing different slopes depending on their individual FSR. The fundamental mode families can be identified by comparing with finite element simulations of the resonator's mode spectrum. Resonances belonging to the $E_{11}^{x}$ ($\mathrm{TE_{11}}$) mode family are underlaid in orange while the $E_{11}^{y}$($\mathrm{TM_{11}}$) family is underlaid in yellow. Modes belonging to the higher order mode families are only partly detected and can not be clearly identified due to their very similar FSRs.

Figure \ref{fig:dispersionCharacterization}c shows the parabolic deviation of the $E_{11}^{y}$ mode family from the central FSR of $95.65\,\mathrm{GHz}$. By fitting with a polynomial the local dispersion of $D_{2}/2\pi=0.49\mathrm{\, MHz}$ is extracted, corresponding to anomalous GVD. This dispersion value is in approximate agreement with the value of $D_{2}/2\pi=\mathrm{0.26\, MHz}$ extracted from simulations (we attribute deviations to inaccuracy in the values of material dispersion). Moreover, avoided modal crossings of the $E_{11}^{y}$ mode family with the $E_{11}^{x}$ family as well as higher order mode families are observed.

\section{Frequency comb generation in the single soliton regime}
\label{sec:Frequency-comb-generation}
\begin{figure}
	\includegraphics[width=1\columnwidth]{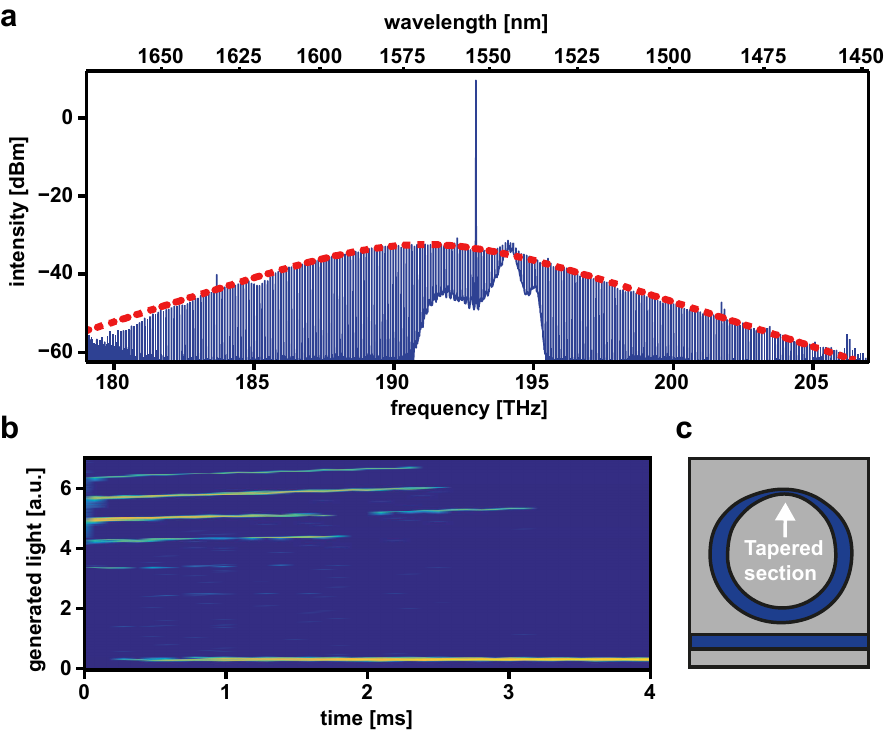}
	
	\protect\caption{\textbf{Frequency comb generation in the single soliton regime. }(a)
		Dissipative Kerr soliton based frequency comb generation
		inside a $95.7\,\mathrm{GHz}$ SiN microresonator pumped with a $6\,\mathrm{W}$
		laser at $192.9\mathrm{\, THz}$. The red line is a fit of the spectral
		envelope with a $\mathrm{sech^{2}}$ function. A $3\,\mathrm{dB}$
		bandwidth of $\mathrm{6.6\, THz}$ is extracted, corresponding to
		a soliton pulse duration of approximately $50\mathrm{\, fs}$. (b)
		Statistics of step distribution in the converted comb light power
		based on 20 consecutive scans of a $3\,\mathrm{W}$ pump laser across
		the cavity resonance. The color-coded histogram reveals millisecond
		timescales for the steps and a higher occurance probability for multi-soliton
		states. (c) Schema of the device layout that comprises a $100\,\mu\mathrm{m}$
		long single mode waveguide section for higher order mode suppression.
		\label{fig:nonlinear_performance}}
\end{figure}

Next we demonstrate nonlinear frequency conversion with our resonators fabricated using the photonic Damascene process. Figure \ref{fig:nonlinear_performance}a shows an optical frequency comb generated upon tuning a $6\mathrm{\, W}$ laser into a resonance of a $95.7\mathrm{\, GHz}$ FSR resonator. The resonator has a radius of $238\,\mu\mathrm{m}$, waveguide width of $1.5\,\mu\mathrm{m}$ and height of $0.9\,\mu\mathrm{m}$. As explained in reference \citep{Kordts2015} the resonator includes a $100\,\mu\mathrm{m}$ long, tapered single mode waveguide section (Figure \ref{fig:nonlinear_performance}c). Within this section the resonator waveguide tapers down to only $0.5\,\mu\mathrm{m}$ width and $0.7\,\mu\mathrm{m}$ height in order to allow only propagation of the fundamental mode families. The device showed nearly critically coupled resonances with $\kappa/2\pi\approx 150\,\mathrm{MHz}$. The single mode waveguide section efficiently suppressed modal crossings due to higher order mode families while not deteriorating the device's quality factor \citep{Kordts2015}.

Upon frequency scanning the pump laser from shorter to longer wavelengths across the resonance, we monitor the converted comb light which includes all comb teeth except the central pump line. Distinct steps in the converted light power, as well as in the total transmitted light power, were observed. Previously such steps have been identified to be a characteristic sign for dissipative Kerr soliton formation inside a microresonator \citep{Herr2013a,Brasch2014}. We obtain statistics of the step formation by repeatedly tuning a $3\,\mathrm{W}$ pump laser across the resonance with approximately $300\,\mathrm{GHz/\mathrm{s}}$ scan rate, while recording the converted light power with an oscilloscope. Figure \ref{fig:nonlinear_performance}b shows a histogram representation (yellow and red denotes higher occurance rate) of 20 consecutive scans. The horizontal lines correspond to different multi-soliton Kerr comb states with more than one pulse circulating in the microresonator: the more pulses circulate, the higher the converted light intensity. Notably, the observed step length in the millisecond time scale is significantly longer then previously observed in SiN microresonators \citep{Brasch2014}, facilitating tuning into the single soliton states.

The frequency comb shown in Figure \ref{fig:nonlinear_performance}a could thus be generated by tuning the pump laser inside the lowest step of the cascade using a laser tuning method, previously applied for crystalline resonators \citep{Herr2013a}. The frequency comb has a smooth spectral $\mathrm{sech^{2}}$ shaped envelope, characteristic for a single temporal dissipative soliton circulating inside the resonator. By fitting with a $\mathrm{sech^{2}}$ envelope function (shown in red) a spectral $3\,\mathrm{dB}$ bandwidth of $\mathrm{6.6\, THz}$ is extracted. The temporal full width at half maximum (FWHM) of dissipative solitons in critically coupled resonators is given as $\Delta t_{\mathrm{min}}^{\mathrm{FWHM}}\approx2\sqrt{\frac{-\beta_{2}}{\gamma\mathcal{F}P_{\mathrm{in}}}}$, where $\mathcal{F}$ is the resonator finesse, $P_{\mathrm{in}}$ the coupled pump power, $\beta_{2}=-(n_{0}/c)\,(D_{2}/D_{1}^{2})$ the group velocity dispersion and $\gamma$ the effective nonlinearity. Based on the experimental parameters (pump power in waveguide $\mathrm{P}=3.28\,\mathrm{W}$, finesse $\mathcal{F}=638$, effective nonlinearity $\gamma=1.41\,\mathrm{W^{-1}m^{-1}}$, dispersion parameter $\beta_{2}=1.95\,\mathrm{fs^{2}m^{-1}}$) we calculate a temporal width of $51.3\,\mathrm{fs}$. This value is in close agreement with the pulse duration of $47.7\mathrm{\, fs}$ calculated for a transform limited $\mathrm{sech^{2}}$ pulse using the measured spectral $3\,\mathrm{dB}$ bandwidth. Furthermore we note the offset between the pump line at $192.9\mathrm{\, THz}$ and the maximum of the hyperbolic secant envelope due to the Raman induced soliton self-frequency shift \citep{Karpov2015}.

\section{Conclusion}

In summary, we have demonstrated a novel photonic Damascene process including a method for efficient thin film stress control. We fabricate SiN microresonators with so far unattainable waveguide dimensions and aspect ratios with close to unity yield. We show Q factors of $3.7\times10^{6}$ - on par with state of the art Q factors obtained using subtractive processes \citep{Barclay2006,Luke2013,Brasch2014} - and achieve broadband temporal dissipative Kerr soliton based frequency comb generation with a $3\,\mathrm{dB}$ bandwidth of $6.6\mathrm{\, THz}$. In the future the high yield and planar top surface of our process will enable integration of nonlinear waveguides with other photonic building blocks e.g. via flip chip bonding integration \citep{Roelkens2010} or novel optoelectronic 2D materials, like graphene or $\mathrm{MoS_{2}}$ \citep{Bonaccorso2010}. Additionally the large waveguide dimensions attainable are required for dispersion engineering of integrated SiN waveguides and microresonator frequency comb generation in the mid infrared spectral region.

\section*{Funding Information}
This publication was supported by Contract W31P4Q-14-C-0050 from the Defense Advanced Research Projects Agency (DARPA), Defense Sciences Office (DSO). This work was further supported by the Switzerland National Science Foundation (SNSF). VB acknowledges the support of the European Space Agency (ESA). MG and MZ acknowledge the support of the Hasler Foundation. JDJ acknowledges the support by a Marie Curie IIF.
\section*{Acknowledgments}
The authors thank the center for MicroNanotechnology CMi at EPFL for technical support.

\bigskip

\bibliography{photonic_damascene_bib}

\end{document}